\documentclass[twocolumn,showpacs,prb]{revtex4}

\usepackage{graphicx}%
\usepackage{dcolumn}
\usepackage{amsmath}

\makeatletter
\def\btt#1{\texttt{\@backslashchar#1}}%
\DeclareRobustCommand\bblash{\btt{\@backslashchar}}%
\makeatother

\begin{document}

\title[Short Title]{Josephson current in $s$-wave superconductor / Sr$_2$RuO$_4$ junctions}

\author{Yasuhiro Asano}
\email{asano@eng.hokudai.ac.jp}
\affiliation{%
Department of Applied Physics, Hokkaido University, 
Sapporo 060-8628, Japan
}%

\author{Yukio Tanaka}
\affiliation{
Department of Applied Physics, Nagoya University, 
Nagoya 464-8603, Japan}%

\author{Manfred Sigrist}
\affiliation{
Theoretische  Physik ETH-H\"{o}nggerberg CH-8093 Z\"{u}rich, 
Switzerland}%

\author{Satoshi Kashiwaya}
\affiliation{
National Institute of Advanced Industrial Science and Technology, 
Tsukuba 305-8568, Japan}%

\date{\today}

\begin{abstract}
The Josephson current between an $s$-wave 
and a spin-triplet superconductor Sr$_2$RuO$_4$ (SRO) is studied
theoretically.  
In spin-singlet / spin-triplet superconductor junctions, 
there is no Josephson current proportional to $\sin \varphi$ in the absence of 
the spin-flip scattering near junction interfaces, where $\varphi$ is 
a phase-difference across junctions. Thus a dominant term of the
Josephson current  
is proportional to $\sin 2\varphi$ . 
The spin-orbit scattering at the interfaces
gives rise to the Josephson current proportional to $\cos\varphi$,
which is a direct 
consequence of the chiral paring symmetry in SRO.
\end{abstract}

\maketitle

\section{Introduction}

The quantum transport through junctions to unconventional superconductors
has attracted 
much attention in recent years, in particular, 
in view of various recently discovered compounds
belonging probably to this class of systems, such 
as Sr$_2$RuO$_4$~\cite{maeno} (SRO),
UGe$_2$~\cite{saxena}, ZrZn$_2$~\cite{pfleiderer}, 
URhGe~\cite{aoki}, CeIn$_3$ and CePd$_2$Si$_2$~\cite{mathur}. 
In such systems zero-energy states~\cite{hu,Buch} (ZES) formed at
interfaces affect crucially the transport properties through
junctions.
In normal-metal / high-$T_c$ superconductor~\cite{bednorz} junctions, 
for instance, a large peak due to the ZES is observed in the
conductance at the zero-bias voltage 
~\cite{tanaka,sk1,sk2,sk3,tanu,wei,iguchi}. 
The resonant tunneling via the ZES enhances
the Andreev reflection~\cite{andreev}, which leads to
the low-temperature anomaly in the Josephson current, e.g. 
between two $d$-wave superconductors
~\cite{tanaka2,tj1,tj2,tj3,tj4,barash,ilichev,asano0}.
The low-temperature anomaly in the Josephson current is a rather
common  
phenomenon for unconventional superconductors including those with
spin-triplet pairing~\cite{asano}. 
The possibility of a logarithmic temperature dependence of the
critical Josephson current was also 
predicted for junctions between two SRO
samples~\cite{barash2,asano2,mahmoodi}.

The Josephson current-phase relation
can be decomposed into a series of contributions
of different order 
\begin{equation}
J = \sum_{n=1}^\infty \left( J_n \sin n\varphi + I_n \cos n\varphi
\right), 
\label{jseries} 
\end{equation}
where $\varphi$ is the phase-difference across junctions.
The coefficients $I_n$ vanish for all $ n $ as long as time reversal
symmetry is conserved, since in this case $ \varphi \to - \varphi $
implies $ J \to - J $. As we will deal in the following 
with a superconducting phase
which break time reversal symmetry, we will keep these
terms. In the most simple approach, the lowest order contribution $J_1$
vanishes, for  a junction of the composition spin-singlet
superconductor / insulator / spin-triplet
superconductor~\cite{geshkenbein,millis,yip,Pals,Fenton,sigrist},
because the 
wave function of the two superconducting condensates are 
in orbital and spin part orthogonal to each other.
In this case the second order contribution with the Josephson current
proportional to $\sin 2\varphi$ is leading. 
The presence of spin-flip scattering and the breaking of
parity at the interface between different materials, would invalidate
this simple-minded argumentation. A magnetically active interface
yielding spin-flip contributions occurs in the presence of spin-orbit
coupling. Obviously, spin-orbit coupling yields new selection rules,
because spin and orbital ``angular momentum'' need not to be conserved
independently, but rather only the ``total angular momentum'' has to
remain unchanged in the tunneling process.
Then lowest order coupling, $J_1$ and/or $I_1$, can be
finite~\cite{geshkenbein,millis,sigrist,asano}, so that 
spin-orbit coupling modifies the current-phase relation of the
Josephson effect between singlet and triplet superconductors qualitatively.  
 
In this paper we study the effect of spin-orbit coupling on the
Josephson effect for the example of the chiral $p$-wave state which is most
likely realized in 
SRO~\cite{laube,rice,mazin,ishida,miyake,sidis,hasegawa,graf,won,matsui,n1,n2,Kuroki,Taki,K,Ogata}.This state breaks time reversal symmetry with
an angular momentum along the c-axis and has inplane equal spin
pairing~\cite{rice}. So far the transport properties in junctions
consisting of SRO and $s$-wave superconductors or  
normal metals have been studied in both theories
~\cite{hasegawa2,yamashiro,yama3,yamashiro2,honerkamp,honer1} and 
experiments~\cite{jin0,jin,liu,sumiyama}. The effect of spin-orbit
coupling, where it had been taken into account, was introduced in form
of effective matrix elements only without the care of a detailed
microscopic model for their origin. Here we will consider a model
which explicitly introduces spin-orbit coupling as an interface effect
and allows us to study the symmetry related issues of the interface
by direct variation of coupling parameters. Our model ignores the
spin-orbit coupling effects in the bulk of the two superconductors for
the reason that details of the ionic lattice and the band structure 
would play an essential role, both of which are not easy to implement
in a simple model of an unconventional superconductor.  
Furthermore, we aim here also at 
effects of the ZES on the Josephson current 
in connection with the spin-orbit coupling.

This paper is organized as follows. In Sec~II, we explain a theoretical model. 
The Josephson current is derived in Sec.~III. The conclusion is given
in Sec.~IV.

\section{Andreev Reflection Coefficients}

We consider here a junction as shown in Fig. 1 between an $s$-wave
superconductor (left hand side) and a $p$-wave superconductor (right
hand side) where the later shall be in the chiral $p$-wave state

\begin{equation}
\boldsymbol{d} (\boldsymbol{p}) \propto  \hat{\boldsymbol{z}} (p_x \pm i p_y
)
\end{equation}
as proposed for SRO. 
The geometry is chosen so that the current flows in
the $x$-direction 
and the $c$-axis of SRO is in the $z$-direction parallel to the
junction interface. Periodic boundary conditions are assumed in the
$y$-direction and the width of the junction is $W$, while the system
is taken homogeneous along the $z$-direction. 
\begin{figure}[htbp]
\begin{center}
\includegraphics[width=9.0cm]{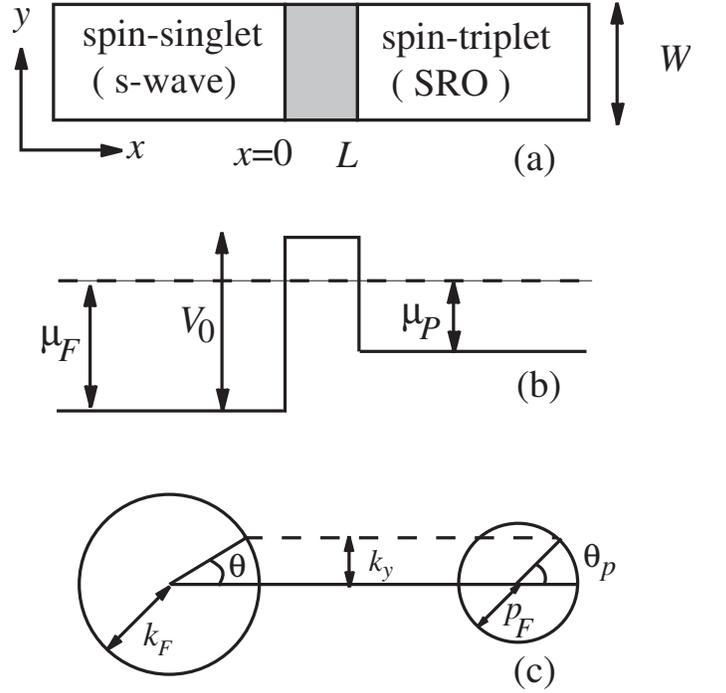}
\end{center}
\caption{
The $s$-wave superconductor / SRO junction is schematically illustrated in (a).
In (b), the broken line indicates the chemical potential of the junction.
In (c), we illustrate the Fermi surface in the two superconductors, where 
$\theta$ and $\theta_p$ are incident angles of a quasiparticle in $s$-wave 
superconductor and in SRO, respectively.
}
\label{system}
\end{figure}
The junction is described by the Bogoliubov-de Gennes (BdG)
equation~\cite{degennes}, 
\begin{align}
\int \!\! d\boldsymbol{r}' &\left[ \begin{array}{cc}
\delta\left( \boldsymbol{r}-\boldsymbol{r}'\right) 
\hat{h}_0(\boldsymbol{r}') & \hat{\Delta}\left(
  \boldsymbol{r},\boldsymbol{r}'\right)\\ 
-\hat{\Delta}^\ast\left( \boldsymbol{r},\boldsymbol{r}'\right) 
& - \delta\left( \boldsymbol{r}-\boldsymbol{r}'\right) 
\hat{h}_0^\ast(\boldsymbol{r}')
\end{array}
\right] 
\left[ \begin{array}{c} \hat{u}(\boldsymbol{r}') \\
\hat{v}(\boldsymbol{r}')\end{array}\right] \nonumber \\
& = E\left[ \begin{array}{c} \hat{u}(\boldsymbol{r}) \\
\hat{v}(\boldsymbol{r})\end{array}\right],
\label{bdg}
\end{align}
\begin{align}
\hat{h}_0(\boldsymbol{r})=& \left[ -\frac{\hbar^2\nabla^2}{2m}-\mu_j + V(\boldsymbol{r})\right]
\hat{\sigma_0}+ \boldsymbol{V}(\boldsymbol{r})\cdot \hat{\boldsymbol{\sigma}}, \label{h0}\\
\hat{\Delta}(\boldsymbol{R},\boldsymbol{r}_r)= &\begin{cases} i 
\boldsymbol{d}(\boldsymbol{r}_r)\cdot \hat{\boldsymbol{\sigma}}\hat{\sigma}_2 & 
\text{for} X_c > L\\
i d_0(\boldsymbol{r}_r) \hat{\sigma}_2 & \text{for}  X_c < 0 
\end{cases},
\end{align}
where $\boldsymbol{R}=(X_c,Y_c)=(\boldsymbol{r}+\boldsymbol{r}')/2$, 
$\boldsymbol{r}_r=\boldsymbol{r}-\boldsymbol{r}'$.  
The unit matrix and the Pauli matrices are denoted as $\hat{\sigma}_0$ and 
$\hat{\sigma}_j$, respectively, with $j=1, 2$ and 3. 
The energy is measured from the chemical potential with 
$\mu =\mu_S$ for $x<L$ and $\mu_P$ for $x>L$,
where $L$ is the  
thickness of the insulator as shown in Fig.\ref{system} (b).
The potential of the insulator is given by
\begin{equation}
V(\boldsymbol{r})= V_0 \left[ \Theta(x) - \Theta(x-L)\right], \label{v01}
\end{equation} 
and is in our model also the source of the spin-orbit scattering described by
the Hamiltonian 
\begin{equation}
H_{so}= -i \left(\frac{\hbar}{2mc}\right)^2 \hat{\boldsymbol{\sigma}} 
\cdot \left[\nabla V_0(\boldsymbol{r}) \times \nabla\right].
\end{equation}
Thus the spin-dependent potential in Eq.~(\ref{h0}) is described as
\begin{align}
\boldsymbol{V}(\boldsymbol{r}) \cdot \hat{\boldsymbol{\sigma}}=& -i
\frac{V_0 \alpha_s}{k_F^2}  
\left[ \delta(x) - \delta(x-L)\right]
\frac{\partial}{\partial y}\hat{\sigma}_3,\label{vspin}\\
\alpha_s =& \left( \frac{\lambda_e k_F}{2}\right)^2,
\end{align}
where $\lambda_e$ is the Compton wavelength and 
$k_F = \sqrt{ 2m \mu_S/\hbar^2}$ is the Fermi 
wave number in the $s$-wave superconductor.
The amplitude of the spin-orbit scattering is characterized by the
dimensionless coupling constant 
$\alpha_s$ which is 
about $10^{-3} \sim 10^{-4}$ in ordinary metals. Throughout this
  paper, $\alpha_s$ is fixed at $10^{-3}$.  
We assume that all potentials are uniform in superconductors.
Therefore the BdG equation can be expressed in the momentum space,
\begin{equation}
\left[ \begin{array}{cc}
\xi_{\boldsymbol{k}} \hat{\sigma}_0 & \hat{\Delta}_{\boldsymbol{k}} \\
-\hat{\Delta}^\ast_{-\boldsymbol{k}} 
& - \xi_{\boldsymbol{k}}\hat{\sigma}_0 
\end{array}
\right] 
\left[ \begin{array}{c} \hat{u}_{\boldsymbol{k}} \\
\hat{v}_{\boldsymbol{k}}\end{array}\right]
= E \left[ \begin{array}{c} \hat{u}_{\boldsymbol{k}} \\
\hat{v}_{\boldsymbol{k}}\end{array}
\right],
\label{k-bdg}
\end{equation}
where we note that $-\hat{\Delta}^\ast_{-\boldsymbol{k}}= \hat{\Delta}^\dagger_{\boldsymbol{k}}$.
In the superconductor with unitary pairing states, 
the amplitudes of the wave function are given by
\begin{equation}
\left[ \begin{array}{c} 
\hat{u}^e_\pm \\
\hat{v}^e_\pm
\end{array}
\right] = 
\left[ \begin{array}{c} 
u_\pm \hat{\sigma}_0 \\
v_\pm \frac{\hat{\Delta}_\pm^\dagger}{|\boldsymbol{D}_\pm |}
\end{array}
\right], \label{wfe}
\end{equation}
in the electron branch and
\begin{equation}
 \left[ \begin{array}{c} 
\hat{u}^h_\pm \\
\hat{v}^h_\pm
\end{array}
\right] = 
\left[ \begin{array}{c} 
v_\pm \frac{\hat{\Delta}_\pm}{|\boldsymbol{D}_\pm |}\\
u_\pm \hat{\sigma}_0 \\
\end{array}
\right], \label{wfh}
\end{equation}
in the hole branch.
In the $s$-wave superconductor, the pair potential and the amplitudes of the wave function 
in Eqs.~(\ref{wfe}) and (\ref{wfh}) are defined by
\begin{align}
\hat{\Delta}_\pm =&  
i \Delta_s \hat{\sigma}_2 \textrm{e}^{i\varphi_s}, \label{dels}\\
u_\pm =& u_s = \sqrt{ \frac{1}{2} \left( 1+
    \frac{\Omega_s}{\omega_n}\right) },\\  
v_\pm =& v_s = \sqrt{ \frac{1}{2} \left( 1-
    \frac{\Omega_s}{\omega_n}\right) },\\  
\Omega_s =& \sqrt{ \omega_n^2+ \Delta_s^2},\\
|\boldsymbol{D}_\pm |=& \Delta_s,
\end{align}
where $\varphi_s$ is a phase of the pair potential in the $s$-wave
superconductor, 
$\omega_n=(2n+1)\pi k_B T$ is the fermionic Matsubara frequency, $k_B$ is the
Boltzmann constant  
and $T$ is a temperature.
For the chiral $p$-wave superconductor, we define now
\begin{equation} 
\boldsymbol{d}(\boldsymbol{k}) = \Delta_p (\tilde{p}_x +i \tilde{p}_y)
\textrm{e}^{i\varphi_p} 
\boldsymbol{z} \quad : (p_x+ip_y-\text{symmtry}),
\end{equation}
where $\varphi_p$ is the order parameter phase,
$\tilde{p}_x = p_x / p_F$, $\tilde{p}_y = p_y / p_F$ and $p_F=\sqrt{2m
  \mu_P/\hbar^2}$  
is the Fermi wavenumber on the right hand side. 
The amplitudes of the wave function in Eqs.~(\ref{wfe}) and (\ref{wfh}) 
are given by
\begin{align}
u_\pm =u_p &= \sqrt{ \frac{1}{2} \left( 1+
    \frac{\Omega_p}{\omega_n}\right) },\\  
v_\pm =u_p &= \sqrt{ \frac{1}{2} \left( 1-
    \frac{\Omega_p}{\omega_n}\right) },\\  
\Omega_p =& \sqrt{ \omega_n^2+ \Delta_p^2},\\
\hat{\Delta}_\pm =& i \boldsymbol{d}_\pm \cdot
    \hat{\boldsymbol{\sigma}} \hat{\sigma}_2,\\ 
\boldsymbol{d}_\pm = & \Delta_p ( \pm \tilde{p}_x + i \tilde{p}_y)  
\textrm{e}^{i\varphi_p}\boldsymbol{z},\label{psro}\\
|\boldsymbol{D}_\pm |=& \Delta_p.\label{delp}
\end{align}
A condition for the formation of the ZES at the surface 
of unconventional superconductors is given by~\cite{asano} 
\begin{equation}
\boldsymbol{d}_- \boldsymbol{d}_+ < 0.\label{zescond}
\end{equation}
In the $p$-wave superconductor, Eq.~(\ref{zescond}) is satisfied only when
a quasiparticle is incident perpendicular to the junction 
interface, (i.e., $\tilde{p}_y=0$). For other momentum directions
subgap states at finite energy appear forming a gapless chiral
quasiparticle spectrum.  
The wave function in the
$s$-wave superconductor 
$\boldsymbol{\Psi}^s(\boldsymbol{r})$ and in the $p$-wave
superconductor
$\boldsymbol{\Psi}^p(\boldsymbol{r})$ can be represented by
Eqs.~(\ref{dels})-(\ref{delp}). 
In the presence of the spin-orbit scattering, as shown in 
Eq.~(\ref{vspin}), the wave functions in the two superconductors
are connected with the wave function in the insulator 
$\boldsymbol{\Psi}^b(\boldsymbol{r})$ via the boundary conditions, 
\begin{align}
\boldsymbol{\Psi}^b(0,y)=&\boldsymbol{\Psi}^s(0,y),\label{c1}\\
\left. \frac{d}{dx}\boldsymbol{\Psi}^b(x,y)\right|_{x=0}=&
\left. \frac{d}{dx}\boldsymbol{\Psi}^s(x,y)\right|_{x=0} \nonumber\\
 &+ \bar{V}_0 \alpha_s k_y \check{S_3}\boldsymbol{\Psi}^s(0,y),\label{c2}\\
\boldsymbol{\Psi}^b(L,y)=&\boldsymbol{\Psi}^s(L,y),\label{c3}\\
\left. \frac{d}{dx}\boldsymbol{\Psi}^b(x,y)\right|_{x=L}=&
\left. \frac{d}{dx}\boldsymbol{\Psi}^p(x,y)\right|_{x=L} \nonumber\\
 &+ \bar{V}_0 \alpha_s k_y \check{S_3}\boldsymbol{\Psi}^p(L,y),\label{c4}\\
\check{S_3}=& \left( \begin{array}{cc} \hat{\sigma}_3&  0\\
                    0 & -\hat{\sigma}_3 \end{array} \right).
\end{align}
Since $\alpha_s$ is a small value, we calculate the Andreev reflection coefficients
within the first order of $\alpha_s$. From Eqs.~(\ref{c1})-(\ref{c4}), 
the Andreev reflection coefficients of a quasiparticle incident from a $s$-wave
superconductor are calculated as 
\begin{align}
\hat{r}^{he}=& \left( \begin{array}{cc} 0 &  r^{he}(\uparrow,\downarrow) \\
                     r^{he}(\downarrow,\uparrow) &  0 \end{array}\right),\\
\hat{r}^{eh}=& \left( \begin{array}{cc} 0 &  r^{eh}(\uparrow,\downarrow) \\
                     r^{eh}(\downarrow,\uparrow) &  0 \end{array}\right),\\
r^{he}(\downarrow,\uparrow) =& \frac{X }{\Xi_+}
\left[ -u_sv_s + u_p \tilde{v}_p  f_1 \right], \\
r^{eh}(\uparrow,\downarrow) =& \frac{X}{\Xi_+}
\left[ -u_sv_s - u_p \tilde{v}_p f_1^\ast \right], \\
r^{he}(\uparrow,\downarrow) =& \frac{X }{\Xi_-}
\left[ u_sv_s + u_p \tilde{v}_p  f_1 \right], \\
r^{eh}(\downarrow,\uparrow) =& \frac{X }{\Xi_-}
\left[ u_sv_s - u_p \tilde{v}_p  f_1^\ast \right], 
\end{align}
with
\begin{align}
\tilde{v}_p =& v_pe^{-i\theta_p}, \\
f_1 =& u_s^2 e^{i\varphi} -v_s^2e^{-i\varphi},\\
X =& 4 kp q^2,\\
\Xi_\pm =& Z_\pm (u_p^2 + \tilde{v}_p^2)(u_s^2-v_s^2)\nonumber\\
+& X \left[ 
 u_p^2u_s^2 - \tilde{v}_p^2v_s^2 -2i u_p \tilde{v}_p u_s v_s
 \sin\varphi\right],\\ 
Z_\pm =& z_0 + X 
\mp \bar{V}_0 \alpha_s q \sin\theta  \sinh( 2 q k_F L)
{\delta\mu},\label{zpm}\\ 
z_0 =& \bar{V}_0 (\bar{V}_0 - \delta\mu) \sinh^2( q k_FL) + (k-p)^2 q^2
\end{align}
where $\varphi= \varphi_s - \varphi_p$, $\bar{V}_0=V_0/\mu_S$ and 
${\delta\mu}=(\mu_S-\mu_P)/\mu_S$.
In what follows, we measure the energy in units of $\mu_S$ and the length 
in units of $1/k_F$.
The wave numbers in the $x$ direction are normalized as
\begin{align}
k=&k_x/k_F=\cos\theta \quad \text{($s$-wave)},\\
q=& q_x /k_F = \sqrt{\bar{V}_0-\cos^2\theta} \quad \text{ (insulator)},\\ 
p=& p_x/k_F=\sqrt{\cos^2\theta-\delta\mu} \quad \text{(SRO)}.
\end{align}
The incident angle of a quasiparticle in the $s$-wave superconductor is
$\theta$ and 
in the $p$-wave superconductor $\theta_p=\arctan(k_y/p_x)$ as
depicted in Fig.\ref{system} (c). 

\section{Josephson Current}

The Josephson current is expressed in terms of the Andreev reflection
coefficients~\cite{asano,nishida}
\begin{align}
J=&\frac{e}{2\hbar}  T\sum_{\omega_n}
\boldsymbol{I}, \\
\boldsymbol{I}=&\frac{N_c}{2} \int_{-\theta_0}^{\theta_0}\!\!\!d\theta
\;\cos\theta \:  
\frac{1}{\Omega_s} \textrm{Tr} 
\left[ \hat{\Delta}_s \hat{r}^{he} - \hat{\Delta}_s^\dagger
  \hat{r}^{eh}\right],  
\label{bi}
\end{align} 
where $\theta_0=\arccos( \delta\mu)$ and 
$N_c=Wk_F/\pi$ is the number of propagating channels on the Fermi surface.
In what follows, we take the units of $\hbar=k_B=1$.
The transmission probability of the junction ($g_J$) is given by 
\begin{align}
g_J=&  \int_0^{\theta_0}\!\! 
d\theta \; \cos\theta \; T_N ,\label{gj}\\
T_N = &\frac{X }{z_0+X},
\end{align}
and $G_J = R_J^{-1}=({2e^2}/{h})N_c g_J$ is the normal conductance of 
the junction.

\begin{figure}[htbp]
\begin{center}
\includegraphics[width=9.0cm]{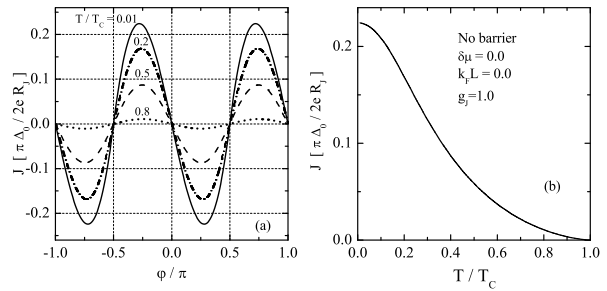}
\end{center}
\caption{
The Josephson current in $s$-wave superconductor / SRO is plotted as a function
of a phase-difference across junctions for several choices of temperatures in (a), where 
$z_0=0.0$ and $\Delta_s =\Delta_p$.
In (b), the maximum Josephson current is plotted as a function of temperatures.
}
\label{z0}
\end{figure}

We first consider the Josephson effect in the absence of the potential
barrier, (i.e., $z_0=0$).
In Fig.~\ref{z0} (a), we show the Josephson current as a function of
$\varphi$,  
where $\delta\mu=0$, $k_FL=0$ and the integration with respect to $\theta$
in Eq.~(\ref{bi}) is carried out numerically. For simplicity we assume 
$\Delta_s=\Delta_p$ and describe the temperature dependence of the
pair potential  
by using the BCS theory. The pair potential at $T=0$ is denoted by $\Delta_0$. 
In this case, $g_J=1$ and there is no spin-orbit scattering at the junctions.
The Josephson current is proportional to $\sin 2\varphi$ for high temperatures
such as $T/T_c = 0.2 \sim 0.8$ in (a). In a very low temperature, $T/T_c=0.01$
proportional to $\sin 4\varphi$ slightly modifies the phase-current
relationship.   
 The maximum amplitude of the Josephson current increases
 monotonically with
 decreasing temperatures 
as shown in Fig.~\ref{z0} (b).

\begin{figure}[htbp]
\begin{center}
\includegraphics[width=9.0cm]{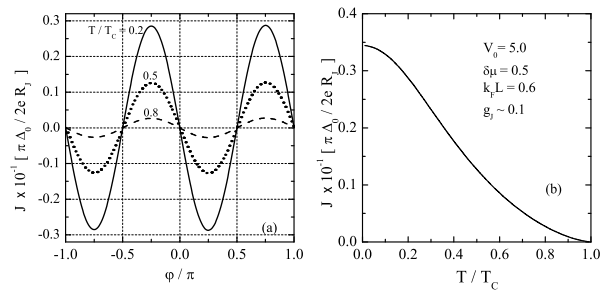}
\end{center}
\caption{
The Josephson current in $s$-wave superconductor / SRO is plotted as a function
of $\varphi$ for several choices of temperatures in (a), 
where $\bar{V}_0=5.0$, $\delta\mu=0.5$ and $ k_FL=0.6$.
In (b), the maximum Josephson current is plotted as a function of temperatures.
}
\label{z1}
\end{figure}

Since electronic structures in $s$-wave superconductor are 
different from those in SRO, complete transparency of the interface is
unrealistic for real junctions. 
Moreover, we consider an insulating layer between the two superconductors.
The calculated results are shown in 
Fig.~\ref{z1}, where $\bar{V}_0=5.0$, $ k_FL=0.6$.
We introduce a finite difference of the chemical potentials 
on both sides of the junction, $\delta\mu=0.5$. This is necessary to
break the symmetry of the junction, otherwise the orbital parts of
different parity would still be orthogonal. 
With these parameters, the transmission probability
is calculated to be $g_J \approx 0.1$.
The phase-current relationship is almost described by $J\propto \sin
2\varphi$ even 
in the presence of the spin-orbit scattering.
Since $\alpha_s$ is the small constant, effects of the spin-orbit scattering 
are still negligible in Fig.~\ref{z1}.
\begin{figure}[htbp]
\begin{center}
\includegraphics[width=9.0cm]{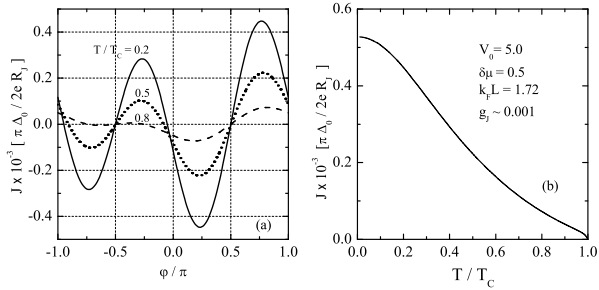}
\end{center}
\caption{
The Josephson current in $s$-wave superconductor / SRO is plotted as a function
of $\varphi$ for several choices of temperatures in (a), 
where $\bar{V}_0=5.0$, $\delta\mu=0.5$ and $ k_FL=1.72$.
In (b), the maximum Josephson current is plotted as a function of temperatures.
}
\label{z5}
\end{figure}

Next we consider the Josephson effect in the limit of $g_J << 1$.
The results in such junctions 
are shown in Fig.~\ref{z5}, where $\bar{V}_0=5.0$ and $\delta\mu=0.5$. 
The thickness of the insulator  $k_FL=1.72$ is much larger than that
in Fig.~\ref{z1} 
and $g_J$ is about 0.001. In Fig.~\ref{z5} (a), the phase current
relationship 
deviates substantially 
from $\sin 2\varphi$ because of the spin-orbit scattering.
At the zero temperature, the Josephson current can be roughly expressed by
\begin{equation}
J \sim - \int_0^{\theta_0} \!\! d\theta\; \cos\theta 
\left[ {\delta\mu\alpha_s}  T_N \cos\varphi
+ T_N^2 \sin 2\varphi\right], \label{jf}
\end{equation}
where $ T_N \sim X/z_0$ because of $z_0 \sim \bar{V}_0^2
\exp(2\sqrt{\bar{V}_0} Lk_F)\gg X$. 
The first term is coming from the spin-orbit
scattering and is proportional to $\cos\varphi$. 
In Eq.(\ref{jseries}), 
$J_1$ and $I_1$ are proportional to $T_N$, therefore they are
proportional to $g_J$  
because a quasiparticle travels twice across the junction
to contribute to $J_1$ or $I_1$.  
A quasiparticle goes across the barrier four times to contribute to 
the Josephson current proportional to $\sin 2\varphi$. Thus the second
term of Eq.~(\ref{jf}) 
is proportional to $T_N^2$.
Generally speaking, $J_n$ and $I_n$ are proportional to $T_N^{n}$ and
$g_J^n$. 
In order to observe the first term in experiments, the transmission
probability of the junction 
must be small enough to satisfy a relation
\begin{equation}
{\delta \mu} \alpha_s \exp(2\sqrt{\bar{V}_0} Lk_F)  \sim 1. \label{so-cond}
\end{equation} 
The left hand side of Eq.~(\ref{so-cond}) is 0.007 in Fig.~\ref{z1}
and 1.1 in  
Fig.~\ref{z5}. To compare with experiments, Eq.~(\ref{so-cond}) should be 
put in another way,
\begin{equation}
\frac{J_c}{(\pi \Delta_0/2eR_J)} \leq \alpha_s \sim 10^{-3},
\end{equation}
where $J_c$ is the critical Josephson current.  
In addition to this, the first term  becomes 
more dominant at temperatures close to $T_c $ as shown in
Fig.~(\ref{z5}) (a).  
 In this  junction, there are ZES states. 
However, there is no anomalous behavior of the Josephson current 
even in low temperatures as shown in Fig.~(\ref{z5}) (b).
As we see shortly this aspect can be attributed to the fact, that it
is the $p_y$ (the transverse component of the pair wave function)
which yields the coupling to the $s$-wave superconductor through
spin-orbit scattering. According to Eq.(\ref{zescond}), however, only
the $p_x$-component generates the ZES, which couples in higher order
only. 
A similar behavior can be found in Josephson effects in 
SRO/I/SRO junctions, where the two superconductors are belonging to
the different 
chirality~\cite{barash2,asano2}.

\begin{figure}[htbp]
\includegraphics[width=7.0cm]{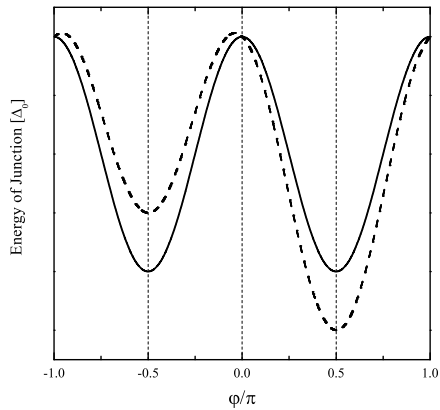}
\caption{
The energy of the junctions is estimated by using the phase-current relationship.
We omit the contribution of the spin-orbit scattering in the solid line. 
The spin-orbit scattering is taken into accout in the broken line. 
}
\label{ene}
\end{figure}

The energy of the junction can 
be calculated from the current-phase relationship by using 
a relation $J = e \partial_\varphi E(\varphi)$. In Fig.~\ref{ene}, we
schematically 
illustrate the energy as a function of $\varphi$.
When the contribution of the spin-orbit scattering is negligible,
there are two energy  
minima at $\varphi=\pm\pi/2$ as shown with the solid line. These
bistable states may be 
used as a base of the quantum computing devices~\cite{ioffe}.  
The spin-orbit scattering breaks the bistability as shown with the
broken line, where 
energy at $\varphi=\pi/2$ is slightly smaller than that at $\varphi = -\pi/2$.
The energy minima do not shift away from $\varphi=\pm\pi/2$ even in the
presence of 
the spin-orbit scattering. Thus, in the absence of the Josephson current,
the phase-difference  
of junction is considerd to be either $\pi/2$ or $-\pi/2$. In experiments, 
the effects of the spin-orbit scattering may be measured from the
difference in the  
critical Josephson current starting from two different energy minima. 
An alternative method to confirm the effects of the spin-orbit scattering is 
the measurement of the Shapiro step in I-V curve. 
It is also possible to observe directly the phase-current
relationship~\cite{ilichev}.

\begin{figure}[htbp]
\includegraphics[width=9.0cm]{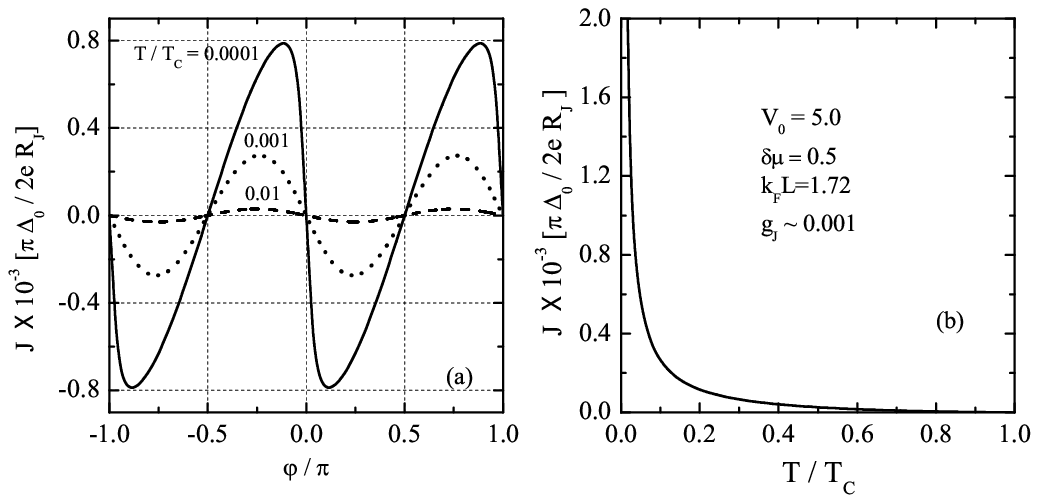}
\caption{
The Josephson current in $s$-wave superconductor / $p_x$-wave superconductor junctions
is plotted as a function
of a $\varphi$ for several choices of temperatures in (a), 
where $\bar{V}_0=5.0$, $\delta\mu=0.5$ and $ k_FL=1.72$.
In (b), the maximum Josephson current is plotted as a function of temperatures.
}
\label{px-fig}
\end{figure}
\begin{figure}[htbp]
\includegraphics[width=9.0cm]{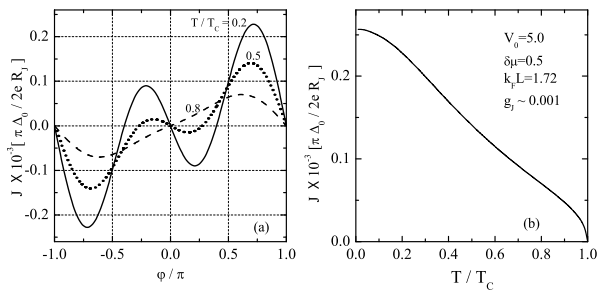}
\caption{
The Josephson current in $s$-wave superconductor / $p_y$-wave superconductor junctions
is plotted as a function
of $\varphi$  for several choices of temperatures in (a), 
where $\bar{V}_0=5.0$, $\delta\mu=0.5$ and $ k_FL=1.72$.
In (b), the maximum Josephson current is plotted as a function of temperatures.
}
\label{py-fig}
\end{figure}

To understand more clearly the relation between the $p_x+ip_y$
symmetry and the phase-current  
relationship, we have also calculated the Josephson current in
$s$-wave superconductor/ 
$p$-wave superconductor junctions with 
other pairing symmetries such as 
\begin{equation}
\boldsymbol{d}_\pm = \pm \Delta_p \tilde{p}_x \textrm{e}^{i\varphi_p}
\boldsymbol{z}  
\quad : (p_x-\text{symmetry})
\label{px},
\end{equation}
and
\begin{equation}
\boldsymbol{d}_\pm = \Delta_p \tilde{p}_y \textrm{e}^{i\varphi_p} \boldsymbol{z}
\quad : (p_y-\text{symmetry}). \label{py}
\end{equation}
In the case of $p_x$-wave symmetry, 
the Josephson current in low transparent junctions is shown 
Fig.~\ref{px-fig}, where $\bar{V}_0=5.0$, $\delta\mu=0.5$ and $ k_FL=1.72$.
The analytical expression at $T=0$ is given by
\begin{equation}
J \sim - \frac{1}{z_0} sgn(\varphi) \cos\varphi.
\end{equation} 
In this case, the spin-orbit scattering gives neither $J_1$ nor
$I_1$. 
At very low temperatures, the Josephson current deviates from $\sin 2\varphi$
because higher harmonics $J_{2n}$ with $n \geq 2$ in Eq.~(\ref{jseries}) 
contributes to the Josephson current as shown in Fig.~\ref{px-fig} (a).
Because the condition in Eq.~(\ref{zescond}) is fulfilled for all 
incident angles of a quasiparticle,
the ZES are formed at the interface. As a consequence, the Josephson current 
shows the low-temperature anomaly and increases 
in proportional to $1/T$ with decreasing temperatures 
as shown in Fig.~\ref{px-fig} (b) \cite{tanaka2,tj1,barash}. \par
In $p_y$-wave symmetry, 
the Josephson current is plotted as a function of $\varphi$ in 
Fig.~\ref{py-fig} (a), where $\bar{V}_0=5.0$, $\delta\mu=0.5$ and $ k_FL=1.72$.
The corresponding analytical result at $T=0$ is given by 
\begin{equation}
J \sim - \int_0^{\theta_0} d\theta\; \cos\theta 
\left[ -\delta\mu\alpha_s  T_N \sin\varphi
+ T_N^2 \sin 2\varphi \right].
\label{jf2}
\end{equation}
In this case, the spin-orbit scattering gives rise to the Josephson
current proportional 
to $\sin \varphi$ which becomes dominant in high temperatures as shown
in Fig.~\ref{py-fig} 
(a). 
Because the condition in Eq.~(\ref{zescond}) is not satisfied,
there is no low-temperature anomaly in the Josephson current as shown in Fig.~\ref{py-fig} (b).

\section{Conclusion}

Finally we summarize the results in this paper.
We calculated the Josephson current in $s$-wave superconductor / SRO
by assuming the pair potential in Eq.~(\ref{psro}) and compare 
the results with another pairing symmetries as shown in
Eqs.~(\ref{px}) and (\ref{py}). 
When the triplet superconductors are described by $p_x$-wave symmetry, 
we find the low-temperature anomaly in the Josephson current because the ZES
are formed at the interface. We also find in $p_x$-wave symmetry that
effects of  
the spin-orbit scattering on the Josephson current are absent. 
It is obvious that spin-orbit coupling is associated with the
transverse component $ p_y $, which does not induce a ZES at the
interface and consequently does not display the anomalous
low-temperature behavior of the maximal Josephson current. 
Thus, the current-phase relation of the junction is connected with the
order parameter phase of the transverse component. This in the case of 
Eq.~(\ref{py}) $ \varphi_p $ and $ \varphi_p + \pi/2 $ for Eq.~(\ref{psro}),
since for the latter the $ p_y $-component is multiplied by $ i $ in
our definition. This finding of the coupling is entirely in agreement
with the qualitative statements of previous discussions of the
Josephson effect of this kind, mediated via spin-orbit
coupling~\cite{geshkenbein,millis,yip,Pals,Fenton,sigrist}.
Furthermore, from Eqs.(\ref{jf},\ref{jf2}) we can conclude, that
beyond the finite spin-orbit coupling represented by the coupling
constant $ \alpha_s $ also the broken parity at the interface, i.e. $
\delta \mu \neq 0 $ is essential for a finite contribution through
this channel.

\end{document}